%% file: main.tex
\documentclass[letterpaper]{article} 
\usepackage{aaai24} 
\usepackage{times}  
\usepackage{helvet}  
\usepackage{courier}  
\usepackage[hyphens]{url}  
\usepackage{graphicx} 
\usepackage{natbib}  
\usepackage{caption} 
\usepackage{algorithm}
\usepackage{algorithmic}
\usepackage{newfloat}
\usepackage{listings}
\DeclareCaptionStyle{ruled}{labelfont=normalfont,labelsep=colon,strut=off} 
\floatstyle{ruled}
\newfloat{listing}{tb}{lst}{}
\floatname{listing}{Listing}
\usepackage[utf8]{inputenc} 
\usepackage[T1]{fontenc}    
\usepackage{url}            
\usepackage{booktabs}       
\usepackage{amsfonts}       
\usepackage{nicefrac}       
\usepackage{microtype}      
\usepackage{color}      
\usepackage{algorithm}
\usepackage{algorithmic}
\usepackage{amsmath}
\usepackage{graphicx} 
\graphicspath{{figure/}}
\usepackage{subfig}
\usepackage{bm}

\newtheorem{theorem}{Theorem}

\DeclareMathOperator*{\argmin}{\arg\!\min}

\newcommand{\optimOper}{\ensuremath{\mathcal{T}^*}}

\newcommand{\qmixSpace}{\ensuremath{\mathcal{Q}^{mix}}}

\newcommand{\qtot}{\ensuremath{Q_{tot}}}

\newcommand{\U}{\ensuremath{\mathbf{u}}}

\newtheorem{prop}{Proposition}

\title{ConcaveQ: Non-Monotonic Value Function Factorization via Concave Representations in Deep Multi-Agent Reinforcement Learning}

 \author{
 \textbf{Huiqun Li}\textsuperscript{\rm 1}, \textbf{Hanhan Zhou}\textsuperscript{\rm 2}, \textbf{Yifei Zou}\textsuperscript{\rm 1}\thanks{Yifei Zou and Tian Lan are corresponding authors.}, \textbf{Dongxiao Yu}\textsuperscript{\rm 1}, \textbf{Tian Lan}\textsuperscript{\rm 2$\ast$}\textsuperscript{\rm}}

\affiliations {
    \textsuperscript{\rm 1} Shandong University 
    \textsuperscript{\rm 2} The George Washington University \\
    hqleee@outlook.com, \{yfzou, dxyu\}@sdu.edu.cn, \{hanhan, tlan\}@gwu.edu
}

\begin{document}

\maketitle

\input{0Abstract}
\input{1Introduction}

\input{2Background}

\input{3Related_works}

\input{4Limitation}

\input{5Method}

\input{6Experiment}

\input{7Conclusion}

\setcounter{section}{0}
\setcounter{theorem}{0}
\setcounter{equation}{0}
\setcounter{prop}{0}

\bibliography{Ref}
\input{8Appendix}
\end{document}

%% file: 0Abstract.tex
\begin{abstract}
Value function factorization has achieved great success in multi-agent reinforcement learning by optimizing joint action-value functions through the maximization of factorized per-agent utilities. To ensure Individual-Global-Maximum property, existing works often focus on value factorization using monotonic functions, which are known to result in restricted representation expressiveness. In this paper, we analyze the limitations of monotonic factorization and present ConcaveQ, a novel non-monotonic value function factorization approach that goes beyond monotonic mixing functions and employs neural network representations of concave mixing functions. 
Leveraging the concave property in factorization, an iterative action selection scheme is developed to obtain optimal joint actions during training. It is used to update agents' local policy networks, enabling fully decentralized execution. 
The effectiveness of the proposed ConcaveQ is validated across scenarios involving multi-agent predator-prey environment and StarCraft II micromanagement tasks. Empirical results exhibit significant improvement of ConcaveQ over state-of-the-art multi-agent reinforcement learning approaches.
\end{abstract}

%% file: 1Introduction.tex
\section{Introduction}
Joint decision-making in multi-agent reinforcement learning (MARL) often requires dealing with 
an exponential expansion of the joint state-action space as the number of agents increases. To this end, recent MARL algorithms~\cite{VDN,NEURIPS2022_c40d1e40,QMIX,chen2023minimizing,chen2022multi} often leverage 
value-function factorization techniques -- which optimize joint action-value functions through the maximization of factorized, per-agent utilities -- to improve learning and sampling efficiency.  

Current value factorization methods mostly adhere to the Individual-Global-Max (IGM) \cite{QMIX} principle, which asserts the coherence between joint and individual greedy action selections. It requires value function factorization to be strictly monotonic, i.e., if local agents maximize their local utilities, 
the joint value function can attain the global maximum. While such monotonic factorization facilitates efficient maximization during training and simplifies the decentralization of the learned policies, it is restricted to {\bf monotonic value function representations} and may lead to poor performance in situations that exhibit non-monotonic characteristics~\cite{WQMIX}. 
In response to the limited representation expressiveness due to monotonic factorization, recent works have considered a number of mitigation strategies, e.g., WQMIX \cite{WQMIX} that formulates a weighted projection toward monotonic functions, QTRAN \cite{QTRAN} that utilizes an additional advantage value estimator, QPLEX \cite{QPLEX} that leverages a dueling structure, and ResQ \cite{ResQ} that introduces an additional residual function network. However, these methods tend to compensate for the representation gap caused by monotonic factorization, subject to the IGM principle.
QTRAN exhibits low sample efficiency and QPLEX may obtain suboptimal results, while WQMIX  tends to possess high approximation errors in non-optimal actions,  and it is difficult for ResQ to find the optimal actions for a scenario requiring highly-coordinated agent exploration~\cite{ResQ}. Going beyond monotonicity and IGM while ensuring decentralized action selection is still an open question\cite{mei2023remix}.

The pivotal insight in this paper is that monotonic value-function factorization may not be necessary for enabling local action selections in MARL. Indeed, we show that monotonic factorization under IGM may lead to suboptimal policies that only recover an arbitrarily small fraction of optimal global actions that are achievable through an ideal policy with full observability.
Instead, we propose a novel approach of concave value-function factorization for high representation expressiveness, called ConcaveQ. It employs a deep neural network representation of concave mixing networks for value-function decomposition. The concave mixing network estimates the joint action-value output $Q_{tot}$ through a concave function $f_{concave}$ of the input local utilities $Q_i$ conditioned on agent $i$'s local observation and action choice, that is, $Q_{tot} = f_{concave}(Q_1, ..., Q_n)$. We propose a deep neural network representation of the concave mixing function $f_{concave}$. It enables a novel concave value-function factorization in MARL.

Due to the lack of IGM under concave mixing functions, the global optimal joint actions are not necessarily consistent with the optimal actions of local agents. ConcaveQ tackles this problem through an iterative action-selection algorithm during training, which attains optimality using the concavity property of the proposed value-function factorization. To support fully decentralized execution, we note that the optimal joint action cannot be obtained directly from maximizing the local agent utilities $Q_i$. ConcaveQ adopts a soft actor-critic structure with local policy networks for distributed execution, such that each local agent can select the best action according to its own local policy network, which are corrected by the concave value networks during training. It also allows auxiliary information such as entropy maximization to be exploited for effective learning.

For evaluation, we demonstrate the performance of ConcaveQ on SMAC maps and predator-prey tasks. We compare ConcaveQ with several state-of-the-art value factorization methods on various MARL datasets. The experimental results show that ConcaveQ outperforms multiple competitive value factorization methods, especially on difficult tasks that require more coordination among the agents since monotonic factorization restricts global value-function estimate and hinders the effective training of agents.
Our numerical results show significant improvement over SOTA baselines in StarCraft II micromanagement challenge tasks regarding higher reward and convergence speed.

%% file: 2Background.tex
\section{Background}

 \textbf{Value Decomposition.} In cooperative multi-agent reinforcement learning, value function decomposition is a paradigm for cooperative multi-agent reinforcement learning (MARL) that aims to learn a centralized state-action value function that can be decomposed into individual agent values. Value decomposition approaches can reduce the complexity and improve the scalability of MARL problems. Two prominent examples of value function decomposition methods are VDN and QMIX, which both assume that the centralized action value $Q_{tot}$ is additive or monotonic with respect to the agent values $Q_i$. VDN simply sums up the agent values to obtain $Q_{tot} = \Sigma_{i=1}^nQ_i(\tau_i, u_i)$, while QMIX uses a mixing network to combine the local agent utilities in a monotonic way. QMIX enforces the monotonicity through a constraint on the relationship between $Q_{tot}(s, \boldsymbol{u}) = f_\theta(s, Q_1(\tau_1, u_1), ..., Q_n(\tau_n, u_n))$ and each $Q_i$, that is, $\frac{\partial{Q_{tot}}(\boldsymbol{\tau}, \boldsymbol{u})}{\partial{Q_i}(\tau_i, u_i)} > 0, \forall i \in \mathcal{N}$. The monotonicity constraint ensures that the optimal joint action for the global reward $Q_{tot}$ is also optimal for each agent’s local utility $Q_i$, where the mixing function $f_\theta$ is formulated as a feed-forward network parameterized by $\theta$. The weights of the mixing network are produced by independent hyper-networks, which take the global state as input and use an absolute activation function to ensure that the mixing network weights are non-negative to enforce the monotonicity. Then QMIX is trained end-to-end to minimize the squared TD error on \textcolor{black}{ mini-batch of $b$ samples from the replay buffer as $\Sigma^b_{i=1}\left(Q_{tot}(\boldsymbol{\tau}, \boldsymbol{u}, s; \theta) - y_{tot} \right)^2$, where $y^{tot} = r + \gamma max_{u'} Q_{tot}(\boldsymbol{\tau'}, \boldsymbol{u'}, s'; \theta')$, $r$ is the global reward and $\theta'$ is the parameters of the target network whose parameters are periodically copied from $\theta$ for training stabilization.}

\noindent {\bf Weighted QMIX Projection.}  QMIX projects $Q_{tot}$ to $\qmixSpace$ by minimizing the projection loss:
\begin{equation}
\argmin_{q \in \qmixSpace} \sum_{\U \in \mathbf{U}} (\optimOper \qtot(s,\U) - q 
(s,\U))^2.
\label{eq:qmix_objective}
\end{equation}
Weighted QMIX \cite{WQMIX} extends the idea of monotonic mixing by introducing a weighting function into the QMIX projection operator to bias the learning process towards the best joint action. The weighted projection can be described as:
\begin{equation}
\mathop{\arg\min}\limits_{q \in Q^{mix}} \sum\limits_{\textbf{u}\in \textbf{U}}(w(s, \textbf{u})Q(s, \textbf{u}) - q(s, \textbf{u}))^2
\end{equation}
where $w$ is the weighting function that is added to place more importance on optimal joint actions, while still anchoring down the value estimates for other joint actions.  Weighted QMIX can be viewed as a projection onto monotonic mixing function space using weighted distance.

%% file: 3Related_works.tex
\section{Related Work}

\noindent {\bf Value Factorization Approaches.}
Value Factorization approaches are widely adopted in value-based MARL  \cite{marl, dop,zhou2022value,gogineni2023accmer,mei2022mac}. 
Current value factorization methods mostly adhere to the
monotonic IGM principle, such as VDN\cite{VDN} and QMIX\cite{QMIX}. VDN \cite{VDN} represents the joint state-action value function $Q_{tot}$ as a sum of per-agent utilities $Q_i$. QMIX\cite{QMIX} employs a mixing network to factorize the $Q_{tot}$ in a monotonic manner.  The monotonicity constraint ensures that the optimal joint action for the global reward $Q_{tot}$ is also optimal for each agent’s local utility $Q_i$. However, these two decomposition methods suffer from structural constraints, limiting the range of joint action-value functions they can effectively represent.

To compensate for the restricted expressiveness of monotonic representation, recent works have explored several mitigation strategies. Specifically, WQMIX \cite{WQMIX} applies a weighted projection to QMIX, which attaches more importance to the optimal joint actions when minimizing training errors. In WQMIX, finding the optimal weight remains an open problem. Furthermore, the approximation errors are high for non-optimal actions. QTRAN \cite{QTRAN}  incorporates an additional advantage value estimator and imposes a set of linear constraints.  QPLEX \cite{QPLEX}  leverages a dueling structure involving value and advantage functions. ResQ \cite{ResQ} masks some state-action value pairs and introduces an additional residual function network. 
Apart from mitigation strategies, there are also various MARL actor-critic methods, such as MADDPG \cite{maddpg}, MAAC \cite{maac}, and COMA \cite{coma}, which use centralized critics and decentralized actors. PAC \cite{pac} decouples individual agents’ policy networks from value function networks to leverage the benefits of assisted value function factorization.   VDAC \cite{VDAC} combined actor-critic structure with QMIX for the joint state-value function estimation. DOP \cite{dop}  employs a network similar to Qatten \cite{qatten} for policy gradients with off-policy tree backup and on-policy TD.

From the previous works mentioned above, we can see a series of works conducted within the confines of the monotonic assumption. However, few of them consider the non-monotonic case. To the best of our knowledge, this paper is the first one considering the MARL in the concave scenario, which is a general case in non-monotonic scenarios.

%% file: 4Limitation.tex
\section{Characterizing Limitations of Monotonic Factorization}
Monotonic factorization restricts the family of global action value functions that can be represented. More precisely, 
$Q_{tot}(\mathbf{s}, \boldsymbol{u}) = f_\theta(\mathbf{s}, Q_1(\tau_1, u_1), ..., Q_n(\tau_n, u_n))$ implies that if $u_1^* = \text{argmax} Q_1(\tau_1, u_1)$ is the optimal action choice for agent 1, $u_1^*$ must also be optimal for all other states with varying $\tau_2,\ldots,\tau_n$. We analyze this restriction and show how it makes monotonic factorization ineffective in representing complex global value functions.

Considering a Markov decision process (MDP) with $n$ agents, where the state of agent $i$ is denoted by $s_i \in S$ and action by $u_i \in A$, for $i = 1, 2, ..., n$, composing a joint state $\textbf{s}$ and joint action $\textbf{u}$. Let $Q_i(s_i, u_i)$ be local agent value utilities, $Q_{jt}(\mathbf{s}, \textbf{u})$ be the unrestricted ground truth of joint action-value function, and $Q_{mono}(\mathbf{s}, \textbf{u})$ be an arbitrary monotonic factorization estimate of $Q_{jt}(\mathbf{s}, \textbf{u})$. Ideally, the monotonic factorization estimate $Q_{mono}$ should be able to recover the exact optimal action selections of $Q_{jt}$. We define $S_{mono}$ as the state set, where $Q_{mono}$ and $Q_{jt}$ have the same optimal joint action, i.e., 
\begin{eqnarray}
S_{mono}=\left\{ \mathbf{s}: \arg\max Q_{mono}(\mathbf{s}, \textbf{u}) = \arg\max  Q_{jt}(\mathbf{s}, \textbf{u}) \right\} \nonumber
\end{eqnarray}
We show that $|S_{mono}|$ can be arbitrarily small compared to the state space $|S|^n$, when the global maximums are uniformly distributed in  $Q_{jt}(\mathbf{s}, \textbf{u})$. Thus, monotonic factorization could only recover an arbitrarily small fraction of action choices.

{
\begin{theorem}
 When $n \geq log_{|S|}(2|A|\cdot log_2|A|) + 1$ and the optimal action choices in $Q_{jt}$ are uniformly distributed, for any monotonic factorization, we have
$\frac{E(|S_{mono}|)}{|S|^n} \leq \frac{e+1}{|A|}\cdot \delta^{n-1}$ for some constant $\delta\in(0,1)$.
\end{theorem}

\noindent\textbf{Proof  Sketch.} We give a sketch of the proof and provide the complete proof in Appendices. Our key idea is to convert the problem into a classic max-load problem~\cite{balls_into_bins}.

\noindent \textit{\textbf{Step1}: Formulate as max-load bin-ball problem.} For each agent $i$ and state $\mathbf{s}$, we consider the optimal action of $Q_{mono}$ as \textit{ball $i$}. Thus, $Q_{jt}$ and $Q_{mono}$ have the same optimal action for agent $i$ if ball $i$ is placed in the bin corresponding to the optimal action of $Q_{jt}$. 


Let $X_i$ denotes that ball $i$ is in bin $i$, that is,  $Q_{jt}$ and $Q_{mono}$ have the same optimal action for  agent $i$, then we have: 
\begin{equation}
    E(|S_{mono}|) = \Sigma_sP(X_1)\cdot P(X_2|X_1) \cdot...\cdot P(X_n|X_{n-1}...X_1) \nonumber
\label{S_mono}
\end{equation}

Define $Y_i$ as the load of bin $i$, that is, the state space such that $Q_{jt}$ and $Q_{mono}$ have the same optimal action for agents except for agent $i$.
Note that the global maximum of $Q_{jt}$ is uniformly distributed over different states, we then analyze $P(X_1)$:
\begin{equation}
P(X_1) =  \frac{E(\Sigma_{s_2...s_n}X_1)}{|S|^{n-1}}
\leq \frac{E(max_{i}Y_i)}{|S|^{n-1}}
\label{P_X_1}
\end{equation}

\noindent \noindent\textit{\textbf{Step2}: Analyze the probability distribution of the load.} According to the Chernoff bound and the Union bound, we have: when $n \geq log_{|S|}(2|A|\cdot log_2|A|) + 1$,
\begin{equation} 
E(max_{i}Y_i) \leq \frac{(e+1) \cdot {|S|^{n-1}}}{|A|}.
\label{maxY}
\end{equation}
where $|A| \geq 1$ is the size of action space. 

Applying Eq. (\ref{maxY}) to Eq. (\ref{P_X_1}) and Eq. (\ref{S_mono}), we have:
\begin{equation}
P(X_1) \leq \frac{e+1}{|A|}
\end{equation}
Using the same argument repeatedly for agents $i=2,\ldots,n$, we can choose $\delta = min_i(P(X_i|X_{i-1}...X_1))$ and $0 < \delta < 1$. Plugging these inequalities into $ E(|S_{mono}|) $, it yields the desired result $\frac{E(|S_{mono}|)}{|S|^n} \leq \frac{e+1}{|A|}\cdot \delta^{n-1}$.

\begin{figure*}[ht!]
\centering
\includegraphics[scale=0.4]{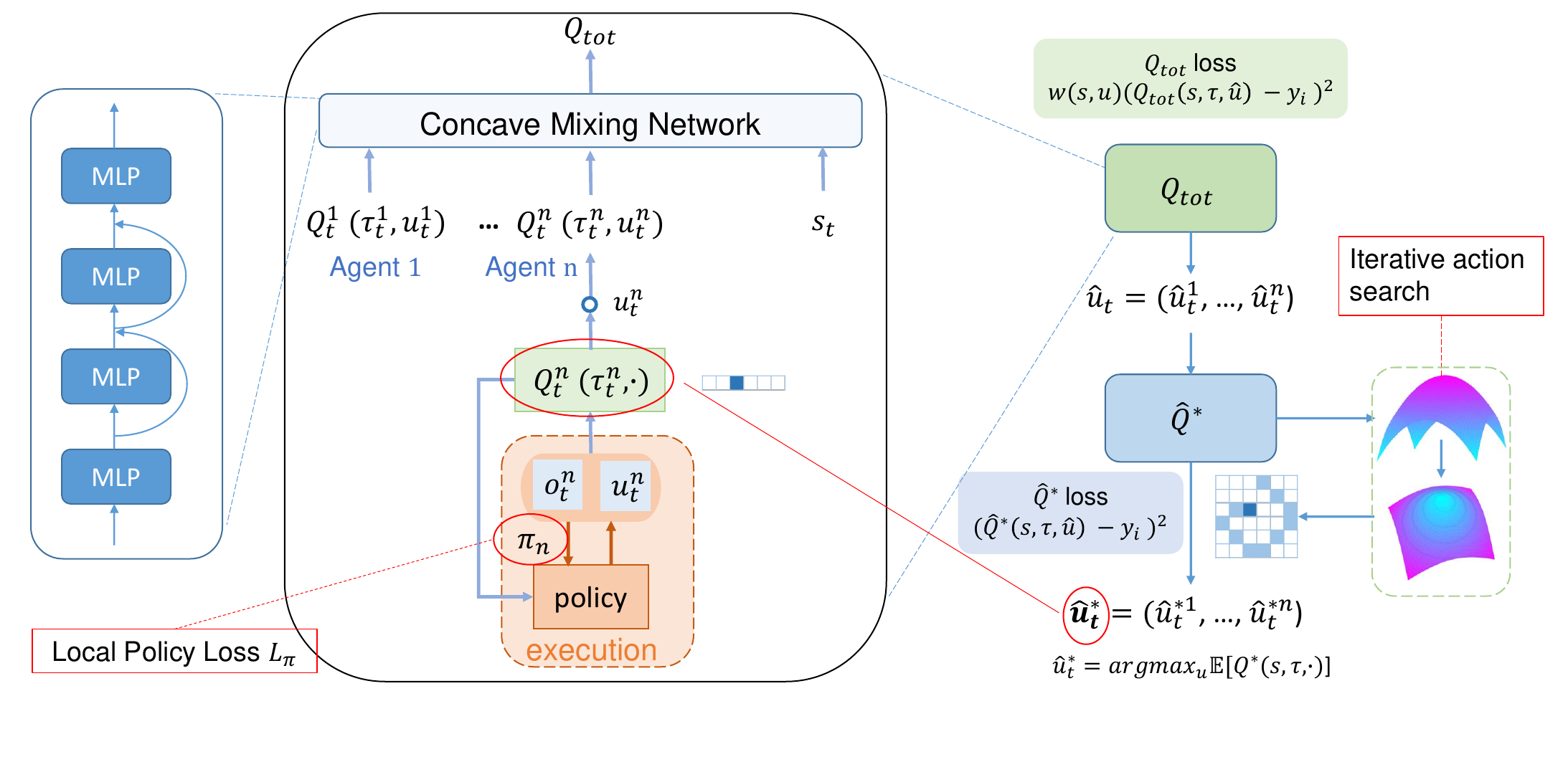}
\caption{The overall architecture of ConcaveQ. The concave mixing network represents $Q_{tot}$ as a concave function of local $Q_a$. With the help of iterative action selection, we can select the optimal actions during training. As for the execution process, the soft policy network is used to find the optimal actions. }
\label{framework}
\end{figure*}

Our analysis shows that a monotonic mixing can only recover an arbitrarily small fraction of global value function maximums, as the number of agents or the size of action space increases. 
Our proposed ConcaveQ addresses the representation issue by introducing a concave mixing network to value function factorization, to enhance the representation ability for better performance. Since the IGM principle no longer holds under concave mixing functions, the greedy action selection by maximizing local agent Q values can not be adopted.
Note that concave optimization has good convergence properties \cite{concaveopt}, that is, a concave function has only one global maximum and the maximum can always be obtained using iterative methods. Therefore,  ConcaveQ tackles the action selection problem by optimizing the joint value function in an iterative manner. As for execution, note that the optimal joint action cannot be obtained directly from maximizing the local agent $Q_i$, we adopt a soft actor-critic policy network in ConcaveQ that uses factorized policies to support distributed execution, such that each local agent can select the best action according to its own local policy network which will be corrected by the Concave value networks during training while exploiting auxiliary information for learning. Moreover, entropy maximization is also used to enhance effective exploration.  As for concave mixing networks, there is always a concave mixing way that can recover the global maximum. 
 
{\begin{prop}
For any state $s$, there is always a concave function $Q_{c}(s, \textbf{u})$ that recovers the global maximum of  $Q_{jt}(s, \textbf{u})$ with the same optimal action.
\end{prop}  

\noindent\textbf{Proof  Sketch.\ } 
For any state $s$, let $f_1(\textbf{u})$ = $Q_{jt}(, \textbf{u})$ and $f_2(\textbf{u})$ = -$Q_{jt}(, \textbf{u})$. According to Fenchel–Moreau theorem \cite{convex_conjugate}, the biconjugate function of $f_2(\textbf{u})$ is a convex function $g(\textbf{u}) = f_2^{**}(\textbf{u})$ and $h(\textbf{u}) = -f_2^{**}(\textbf{u})$  is a concave function. $g(\textbf{u})$ and $h(\textbf{u})$ satisfies:
 \begin{equation}
    g(\textbf{u}) \leq f_2(\textbf{u}), 
\end{equation}
 \begin{equation}
    h(\textbf{u}) \geq f_1(\textbf{u}).
\end{equation}
Suppose the optimal joint action of $f_1(\textbf{u})$ and $h(\textbf{u})$ are $\textbf{u}_{f_1}^*$ and $\textbf{u}_h^*$ respectively,  if we shift $h(\textbf{u})$ by $(-\textbf{u}_{f_1}^* + \textbf{u}_h^*)$, the shifted concave function has the same optimal action as $f_1(\textbf{u})$. In other words, for any state $s$, there is always a concave function that has the same optimal action as $Q_{jt}(s, \textbf{u})$. 
}

The key to our method is the insight that it is not necessary to use the monotonic factorization of QMIX  to extract decentralized policies that are fully consistent with their centralized counterpart. Next, we will discuss concave mixing network design and our ConcaveQ framework to support fully decentralized execution.

%% file: 5Method.tex
\section{Method}
{Since concave factorization may not satisfy the IGM principle, we introduce an iterative action selection algorithm to ensure optimal joint action selection during training. It iteratively optimizes each agent's action selection with all other agent's actions and local utilities fixed. Due to the concave property of the mixing function, the iterative algorithm converges to global optimal actions. 
Note that this iterative action selection does not directly allow decentralized execution. To this end, we further leverage the soft actor-critic framework to train local factorized policies for distributed execution.
Figure \ref{framework} shows the architecture of our learning framework. There are four main
components in ConcaveQ : (1) a concave $Q_{tot}$ utilizing per-agent local utilities, where a concave function serves as the mixing network (2) an unrestricted joint action estimator $\hat{Q}^*$,
which serves as a baseline estimator of the true optimal value function $Q^*$ (3) an iterative action-selection module to seek the optimal joint action during training, and (4) a soft-policy network that allows for the utilization of a soft actor-critic network that consists of local policy networks and local value networks to perform decentralized execution.

\subsection{Concave Mixing Network Design} 

The concave mixing network constrains the output of the network to be a concave function of the inputs, which covers a large class of functions and allows efficient inference via optimization over inputs to the network. Moreover, the concave feature ensures that maximizing the output of the network leads to only one solution. This means that theoretically there won't be local maximum or multiple solutions that may confuse the inference process. Concave mixing network has a suitable property specifically for value function factorization-based multi-agent reinforcement learning problems, that is, local maximization equals global maximization and it can be obtained through iterative methods, which is of great use in our case.

 We first consider a $k$-layer, fully connected network as shown in the left part of Figure \ref{framework}. A concave neural network is defined over the input $x$ for $i = 0, ...,k - 1$ using such architecture:
\setlength\intextsep{0pt}

\begin{equation}
\begin{array}{l}
\begin{aligned}
z_{i+1} = \left\{
\begin{aligned}
&W^{(z)}_0x + b_0, i = 0 \\
&r_i(W^{(z)}_i[z_1, ..., z_i]  + b_i), i = 1, ..., k - 2\\
&-W^{(z)}_i[z_1, ..., z_i]  + b_i, i = k - 1
\end{aligned}
\right.  
\end{aligned}
\end{array}
\end{equation}

where $z_i$ denotes the layer activations, $r_i$ are non-linear activation functions, and $\theta = (W^{(z)}_{0:k-1}, b_{0:k-1})$ are the parameters of linear layers. We can model the concave mixing network as:
\begin{equation}
    f(x;\theta) = z_k
\end{equation}

{
\begin{theorem}
The mixing network $f$ is concave in $x$ provided that all $W^{(z)}_{1:k-1}$ are non-negative, and all functions $r_i$ are convex and non-decreasing.
\end{theorem}
}
 We delineate the proof and the detailed proof is provided in Appendices. The proof is simple and follows from the fact that non-negative sums of convex functions are also convex and that the composition of a convex and convex non-decreasing function is also convex, and that the negative of a convex function is a concave one \cite{convex_optimization}. In our framework, we satisfy the constraint that the $r_i$ should be convex non-decreasing via nonlinear activation units, specifically we adopt the rectified linear functions. The constraint that the $W^{(z)}$ terms be non-negative is somewhat restrictive, but because the bias terms can be negative, the network still has substantial representation power. In our multi-agent reinforcement learning setting, we model $Q_{tot}(s, \textbf{u}; \theta)$ using an input-concave mixing function and select actions under the concave optimization problem $u^*(s) = \text{argmax}_u{Q_{tot}(s, \textbf{u}; \theta)}$.  We have 4 layers, $k = 4$ and the weights  $W_{0:k-1}^{(z)}$ are generated by hypernetworks. Each hypernetwork takes the state
$s$ as input and consists of a single linear layer, succeeded by an absolute ReLU activation function, to generate the non-negative weights.

\subsection{Iterative Action-Selection and Local Policies} We note that the optimal joint action of $Q_{tot}$ cannot be obtained directly from maximizing $Q_i$ since the mixing function is concave rather than monotonic. Hence it is almost impossible to directly find the maximum $Q_{tot}$ and the corresponding joint action selections $\textbf{u}^*$ such that $\textbf{u}^{*} = \text{argmax}_{\textbf{u}} \mathbb{E}[Q(s, \textbf{u})]$, as this takes $O(|U|^{|n|})$ iterations to find the maximum. One of the key insights underlying this method is that although it is impractical to find  ${\textbf{u}^*}$ directly: however, its local estimation and local optimum $\textbf{u}^* = \text{argmax}_\textbf{u}\mathbb{E}[Q^*(s, \tau, \cdot)]$, which takes $O(|U|*{|n|})$ time, is possible to find which will update the concave mixing network and local policy network asymptotically.

To illustrate this problem, we consider the comparison between the action selection from the monotonic mixing network and concave mixing network: in value-based methods with a monotonic mixing network, each agent chooses $u^i_t = \text{argmax}(q^i(\tau^i_t))$, i.e. each agent chooses the action with the best local value thus getting an optimal global value; however, when the utilizing a concave mixing network for a more general value function factorization, we can maximize $\mathbb{E}[Q^*(s, \tau, \cdot)]$ via iterative action selection.

Next, since the mixing network is non-monotonic (concave), we can no longer follow the IGM principle to adopt actions that maximize individual values from the local value networks. Instead, we adopt a soft actor-critic policy network that uses factorized policies to enable distributed execution, such that each local agent can choose the best action according to its local policy network which will be corrected by the concave value networks during training while exploiting auxiliary information for learning.  Under a soft-actor-critic paradigm, each agent can choose $u^i_t = \text{argmax} (\pi^i(\tau^i_t))$ such that  $\mathbb{E}[Q^*(s, \tau, \textbf{u})] = \text{max}(\mathbb{E}[Q^*(s, \tau, \cdot)])$. This ensures that the concave mixing network could factorize concave value functions while each local agent could choose their best actions based on local policy in a decentralized manner during execution. Previous studies have demonstrated that Boltzmann exploration policy iteration can guarantee policy improvement and optimal convergence with infinite iterations and complete policy evaluation. With factorized policy, we have local policy trained with the following: 
\begin{small}
\begin{equation}
\begin{array}{l}
\begin{aligned}
\mathcal{L}_{\pi}(\theta) &=\mathbb{E}_{\mathcal{D}}\left[\boldsymbol\alpha \log \boldsymbol{\pi}\left(\boldsymbol{u}_{t} | \boldsymbol{\tau}_{t}\right)-Q^{\pi}_{tot}\left(\boldsymbol{s_{t}}, \boldsymbol{\tau_{t}}, \boldsymbol{u}_{t}\right)\right] \\
&= -q^{\pi}\left(\boldsymbol{s}_{t}, \mathbb{E}_{\pi^{i}}\left[\text{q}^{i}\left(\tau_{t}^{i}, u_{t}^{i}\right)-\alpha^{i} \log \pi^{i}\left(u_{t}^{i} | \tau_{t}^{i}\right)\right]\right)
\end{aligned}
\end{array}
\end{equation}
\end{small}
where $\text{q}^{\pi}$ is the local value network with $u_i \sim \pi_i(o_i)$, $\mathcal{D}$ is the replay buffer for sampling training data, and the parameter $\alpha$ from soft-actor-critic controls how much the agents prefer actions with higher expected rewards over actions with more explorations. We introduce the training of the local value network and other components in the following sections. The algorithm has a more vital exploration ability and a higher level of generalization.

\begin{figure*}[ht!]%
    \centering
    \subfloat[Punishment $p=0.0$]{
        \includegraphics[width=0.5\linewidth]{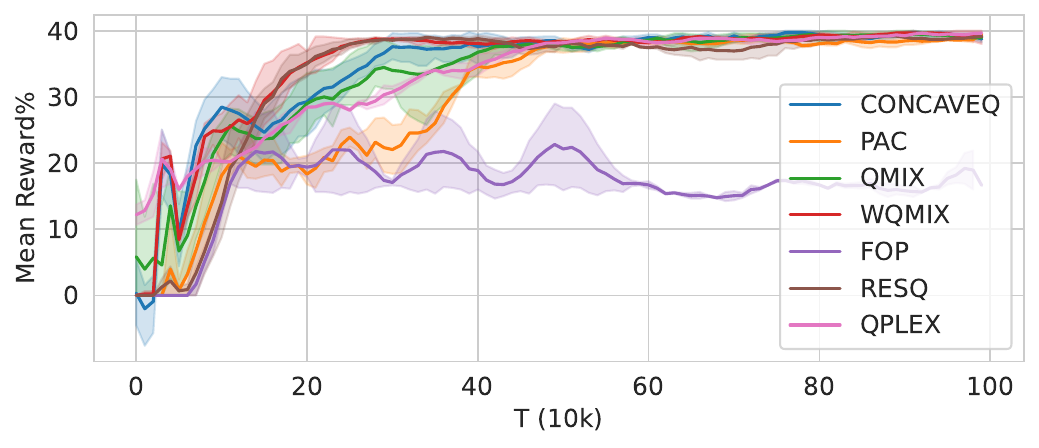}
        }
    \subfloat[Punishment $p = -0.5$]{
        \includegraphics[width=0.5\linewidth]{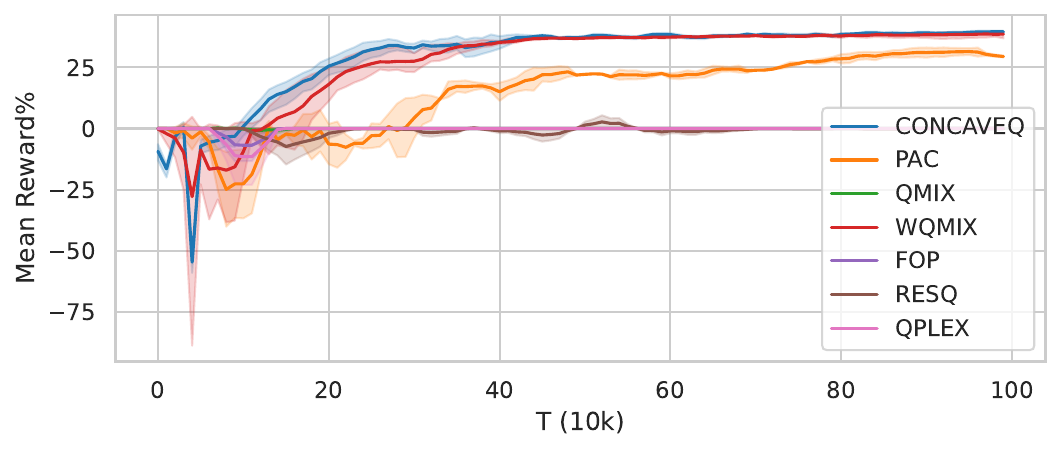}
        }\\
    \subfloat[Punishment $p = -1.5$]{
        \includegraphics[width=0.5\linewidth]{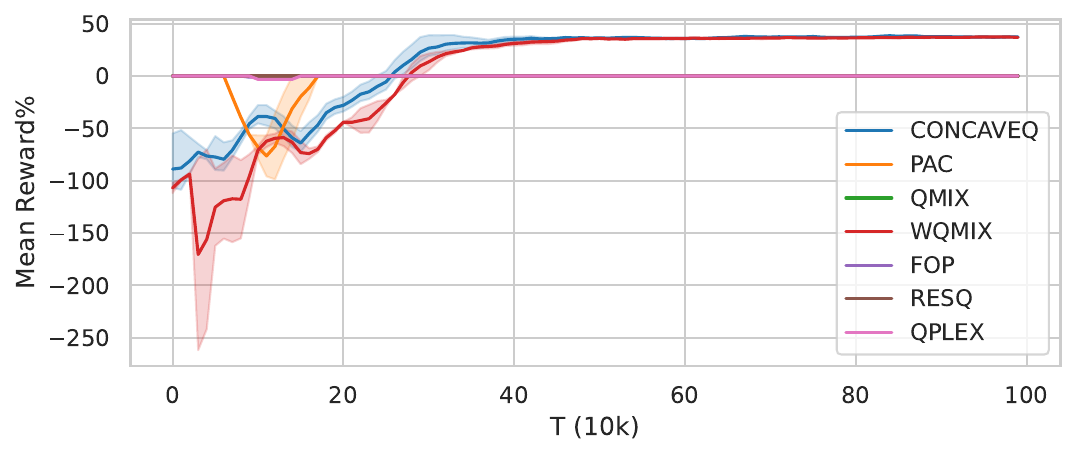}
        }
    \subfloat[Punishment $p = -2.0$]{
        \includegraphics[width=0.5\linewidth]{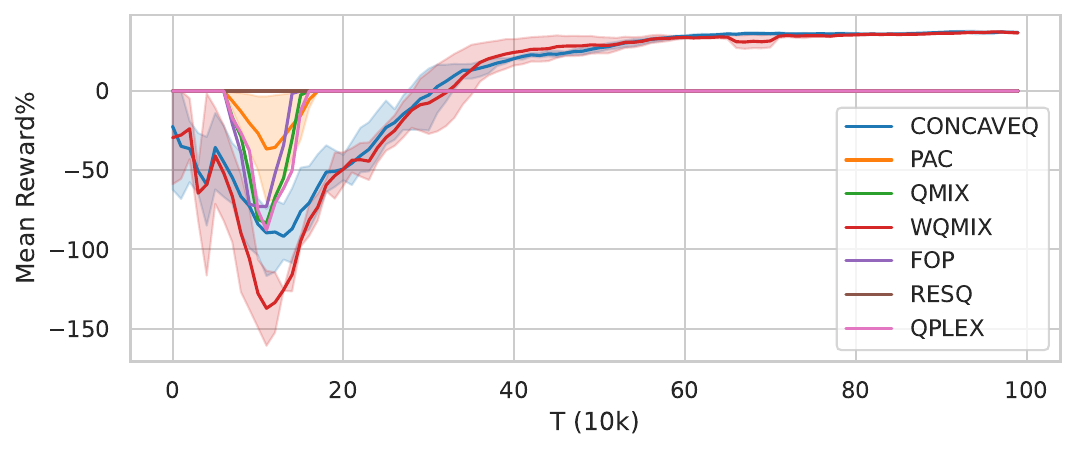}
        }\\    
    \caption{Average reward on the Predator-Prey tasks.}
    \label{exp_stag_hunt}
\end{figure*}

\subsection{Training ConcaveQ} 

We have presented the components of our approach so far. Next, we describe how to implement and train our novel RL algorithm that scales well under DEC-POMDP. During the decentralized execution phase, only the agent network (local policy in Fig.1) is active with actions chosen greedily from each local policy network as $u^i_t = \text{argmax} (\pi^i(\tau^i_t))$, ensuring full CTDE.

The $\hat{Q}^*$ architecture (Green part in Fig. 1) is used as the estimator for ${Q}^*$ from unrestricted functions, where it serves as a mixing network using a feed-forward network that takes its local utilities. Together with the proposed concave mixing network, they are trained to reduce the following loss:

\begin{equation}
\mathcal{L}_{\hat{Q^{*}}}({\theta}) = \sum_{i}
(\hat{Q^{*}}(s, \boldsymbol{\tau}, \hat{\boldsymbol{u}}) - y_i)^2
\label{eq:qcentral_loss}
\end{equation}

\begin{equation}
\mathcal{L}_{{\text{ConcaveQ}}}({\theta}) = \sum_{i}
w(s,\textbf{u})(Q_{tot}(s, \boldsymbol{\tau},\textbf{u}) - y_{i})^2 
\end{equation}

\noindent where $\hat{\boldsymbol{u}} = \operatorname{argmax}_{\boldsymbol{\hat{u}}} Q_{tot}(\boldsymbol{\tau^{\prime}}, \boldsymbol{\hat{u}^{\prime}}, s^{\prime}; \boldsymbol{\theta})$ is from local iterative action selection, $y_{i}=r+\gamma \hat{Q}^{*}(s', \boldsymbol{\tau^{\prime}}, \boldsymbol{\hat{u}})$ and 
$\boldsymbol{\theta^{\prime}}\ $ is the parameters of the target network that are periodically updated to stabilize the training.  $w(s,\boldsymbol{u})$ is the weighting function with $w = 1$ if $Q_{tot}(s, \tau, \boldsymbol{u}) - y_{i} < 0$,  $w = 0.5$ otherwise as suggested in WQMIX \cite{WQMIX}.  Then all components are trained in an end-to-end manner as:

\begin{equation}
\mathcal{L}(\theta) = \mathcal{L}_{\pi} +  \mathcal{L}_{\hat{Q^{*}}} + \mathcal{L}_{{\text{ConcaveQ}}}
\end{equation}

We provide detailed derivations and pseudo-codes for the training process in the appendices.

%% file: 6Experiment.tex
\section{Experiment Results}

\begin{figure*}[ht!]%
    \centering
    \subfloat[3s\_vs\_5z (hard)]{
        \includegraphics[width=0.33\linewidth]{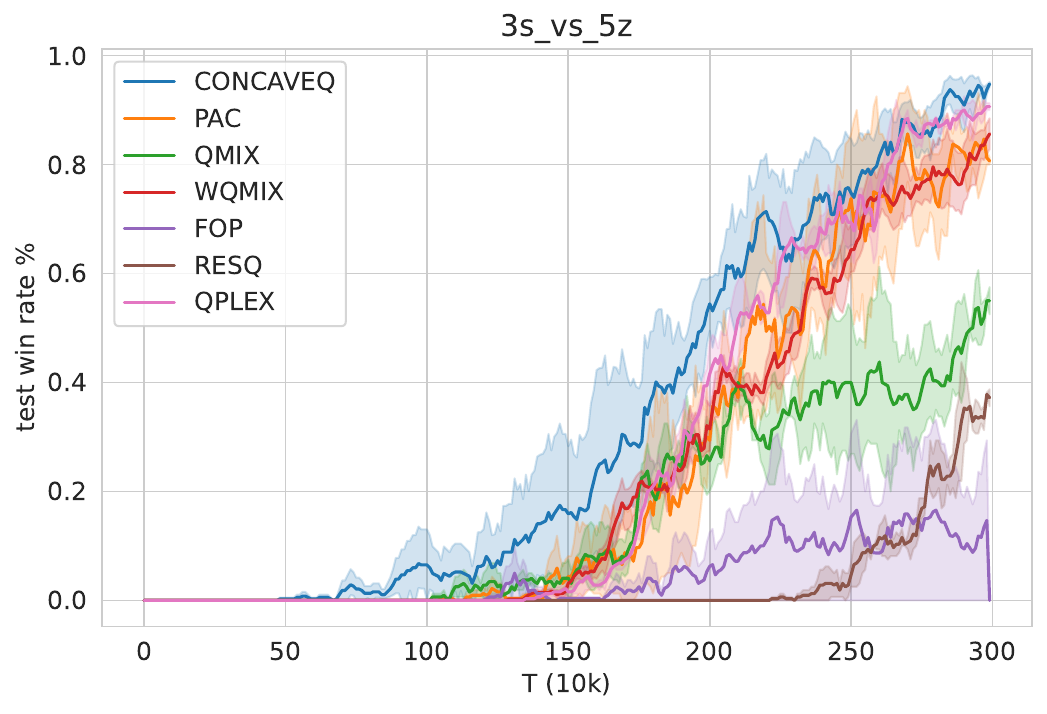}
        }
    \subfloat[5m\_vs\_6m (hard)]{
        \includegraphics[width=0.33\linewidth]{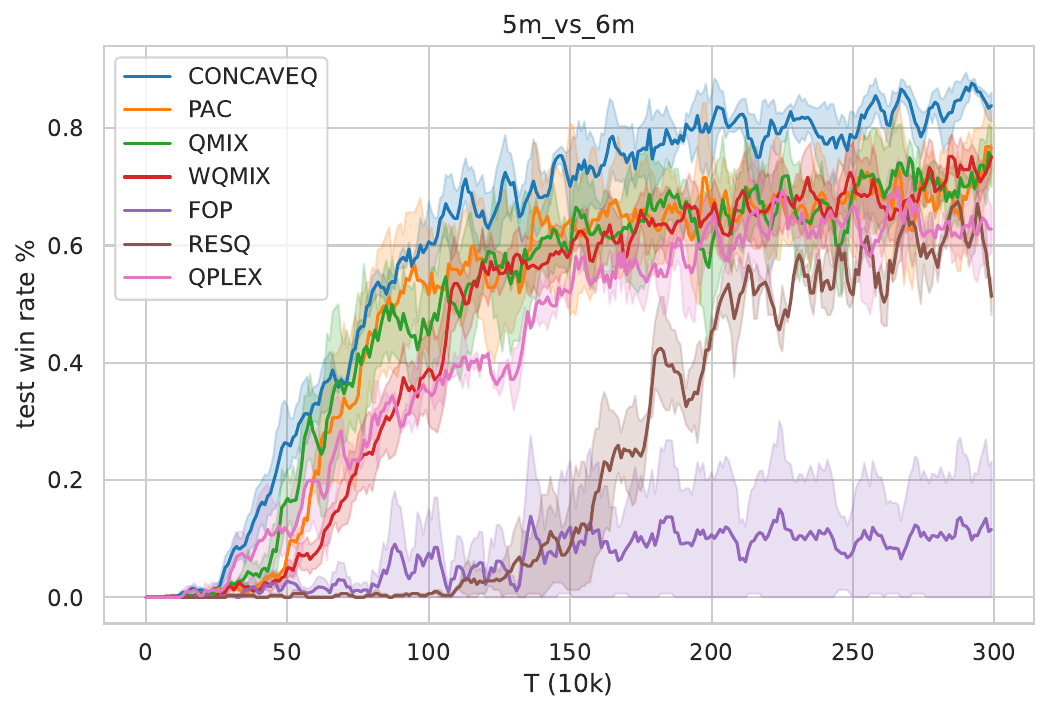}
        }
    \subfloat[27m\_vs\_30m (Super hard)]{
        \includegraphics[width=0.33\linewidth]{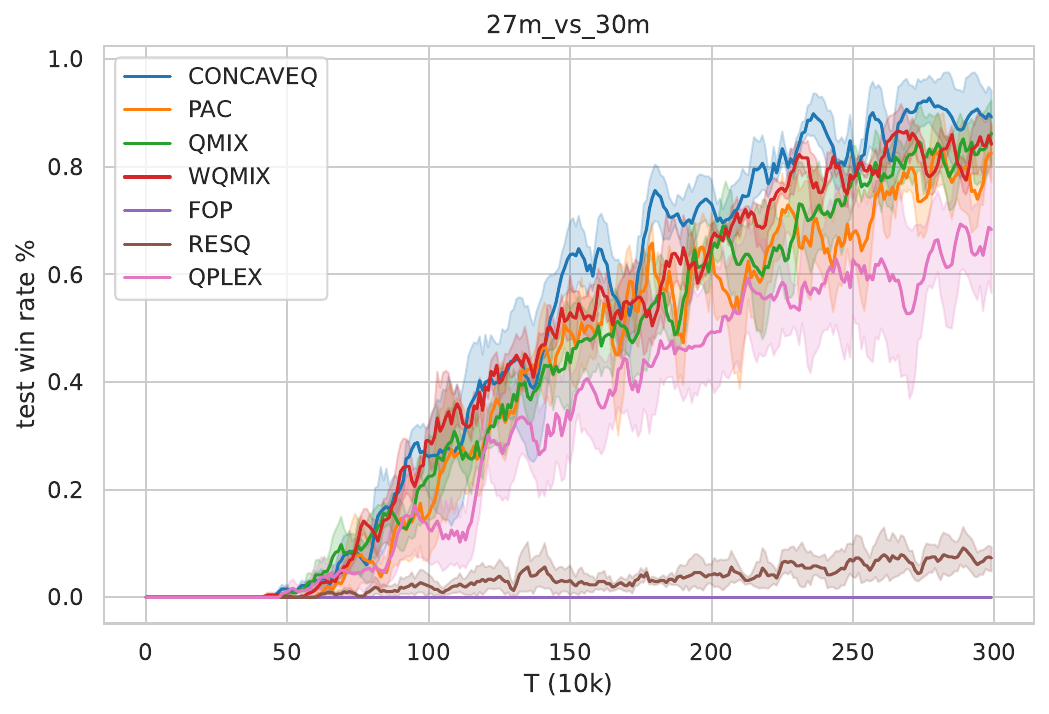}
        }\\
        \subfloat[6h\_vs\_8z (Super hard)]{
        \includegraphics[width=0.33\linewidth]{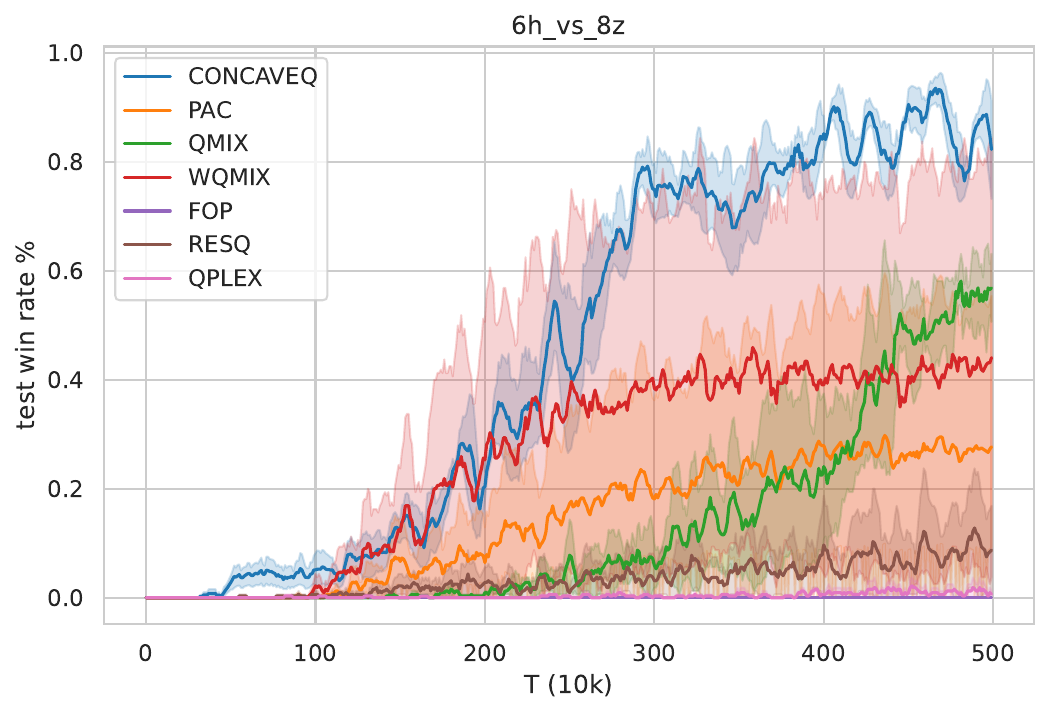}
        }
        \subfloat[Corridor (Super hard)]{
        \includegraphics[width=0.33\linewidth]{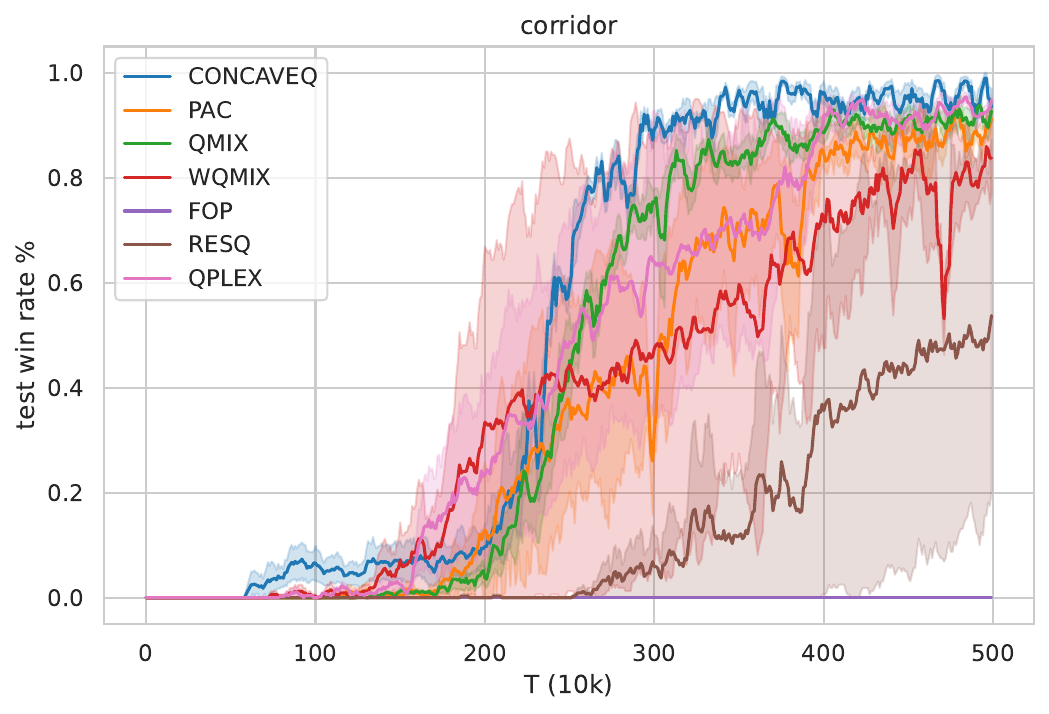}
        }
    \subfloat[MMM2 (Super hard)]{
        \includegraphics[width=0.33\linewidth]{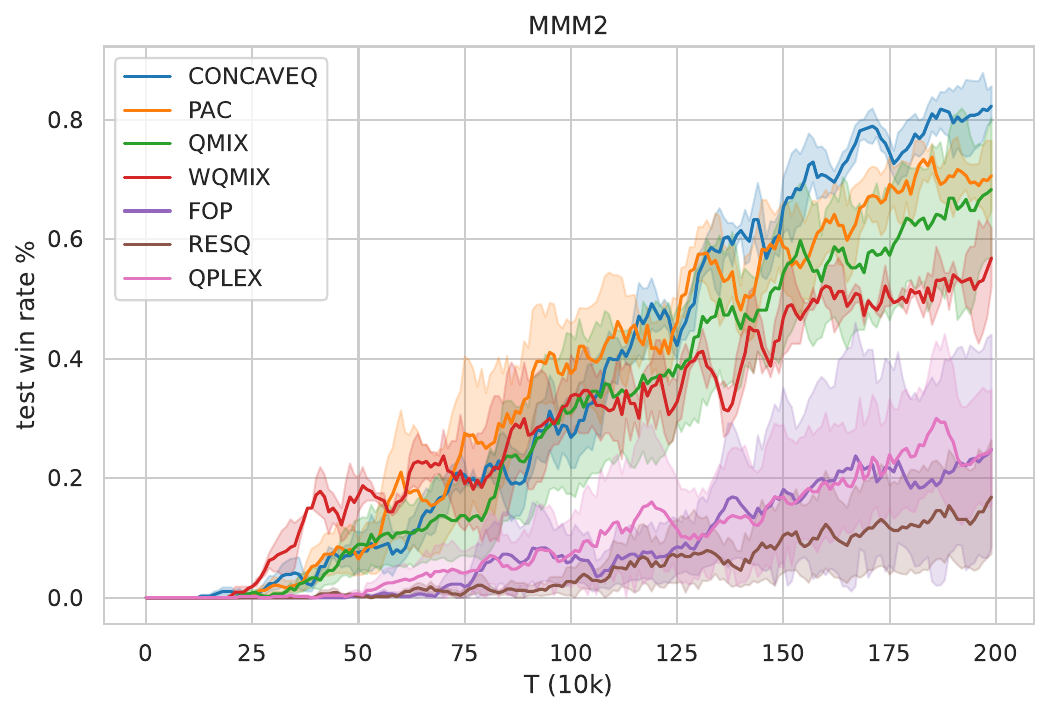}
        }\\    
    \caption{Average test win rate on the SMAC tasks.}
    \label{exp_smac}
\end{figure*}
In this section, we present our experimental results with the state-of-the-art methods on Predator-Prey and the StarCraft II Multi-Agent Challenge (SMAC)\cite{samvelyan2019starcraft}. For fair evaluations, the hyper-parameters of all algorithms under comparison as well as the optimizers are the same. Each experiment is repeated 3 times with different seeds. The presented curves are smoothed by a moving average filter with its window size set to 5 for better visualization. More implementation details and experimental introduction and settings can be found in the Appendices.

\subsection{Predator Prey}
Predator-Prey is a complex partially observable multi-agent environment, where  8 agents cooperate as predators to hunt 8 prey within a  $10 \times 10$ grid. If two or more adjacent predator agents carry out the \textit{catch} action simultaneously, it is a  successful catch and the agents receive a reward $r = 10$. A failed attempt where only one agent captures the prey will receive a punishment reward $p \leq 0$. The more negative $p$ is, the higher level of coordination is needed for the agents.

We select multiple state-of-the-art MARL approaches as baseline algorithms for comparison, which includes value-based factorization algorithms (i.e., QMIX in \cite{QMIX}, WQMIX in \cite{WQMIX}, PAC in \cite{pac},  QPLEX in \cite{QPLEX}, and RESQ in \cite{ResQ}), and decomposed actor-critic approaches (i.e., FOP \cite{fop} ).  
Fig~\ref{exp_stag_hunt} shows the performance of our scheme and the six baselines when punishment $p$ varies from $0$ to $-2$, in which the $x$-axes and $y$-axes represent the number of training episodes and the test mean rewards, respectively. According to the curves in Fig~\ref{exp_stag_hunt}, the following results can be observed. (1) When $p=0$,  CONCAVEQ, PAC, WQMIX, and QMIX can learn good policies and obtain the highest reward when they have been trained for more than $50$ episodes. In other words, the performance of our algorithm is as good as the state-of-the-art works. (2) When punishment gets larger, i.e. $p=-0.5, p= -1.5$ and $p=-2.0$, only CONCAVEQ and WQMIX can still achieve high rewards within $60$ episodes while CONCAVEQ is able to converge faster than WQMIX, the others gradually fail due to their monotonicity constraints or representational limits. These results demonstrate CONCAVEQ's ability in challenging cooperative MARL tasks that require non-monotonic action selections.

\subsection{StarCraft II Multi-Agent Challenge (SMAC)}
In SMAC, agents are divided into two teams to cooperate with allies and compete against enemies or against the other team controlled by the built-in game AI. In the simulation, the agents act according to their local observations and learning experiences.

Note that the state-of-the-art algorithms have already achieved a very good performance on the easy and medium maps, which makes it difficult to present clear comparisons and potential improvements. We carry out our experiment in six maps consisting of two hard maps (\texttt{3s\_vs\_5z, 5m\_vs\_6m}) and four super-hard maps (\texttt{ 27m\_vs\_30m, 6h\_vs\_8z, MMM2, corridor}). The baseline algorithms are the same as those in the Predator-Prey environment.

Details of the environment setting and other training details like network hyperparameters can be found in Appendices.

The performance of our algorithm and the baselines in those hard and super hard maps are presented in Figure \ref{exp_smac}, in which the $x$-axes and $y$-axes represent the number of training episodes and the test win rate, respectively. For almost all scenarios, we found that our algorithm is able to converge faster and deliver a higher win rate. Especially on the \texttt{6h\_vs\_8z} map CONCAVEQ significantly outperforms other baselines by a large margin. We note that the \texttt{6h\_vs\_8z} map requires a sophisticated strategy to take the win, where one unit will be drawing firepower from the enemy units while circling around the edge of the map so the rest friendly units can deliver damages. This implies CONCAVEQ imposes a higher capability of exploration and function representational abilities.

\subsection{Ablation Studies}
\begin{figure}
\centering
\includegraphics[width=0.45\textwidth]{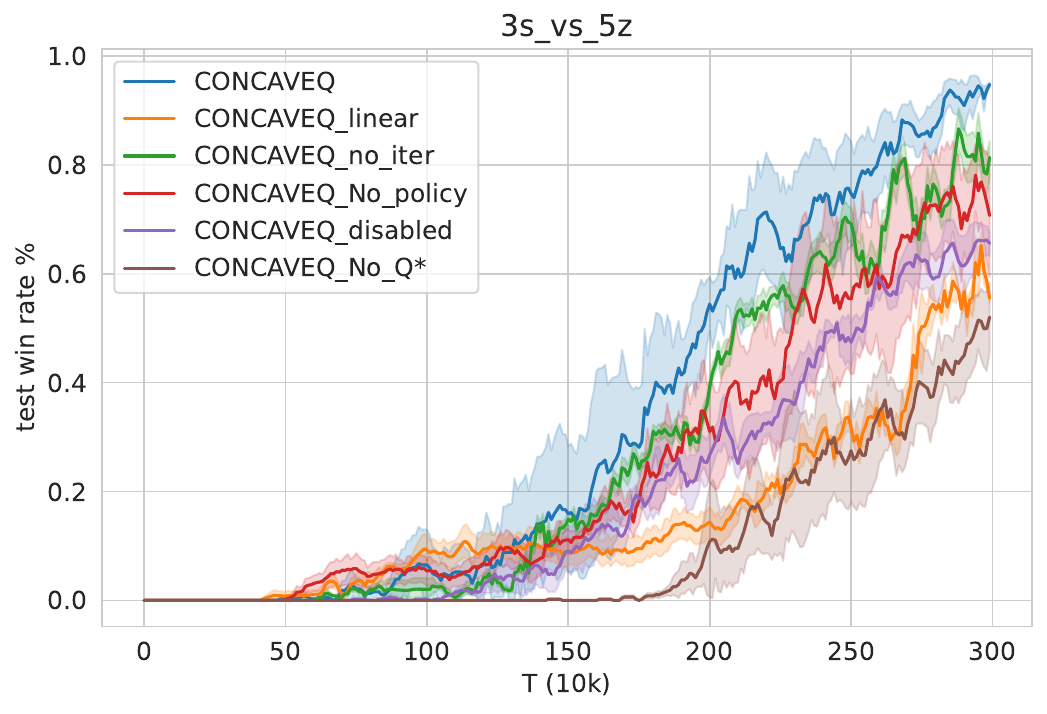}
\caption{Ablations results comparing CONCAVEQ and its  ablated versions on SMAC map \texttt{3s\_vs\_5z}}
\label{exp_ablation}
\end{figure}

We carry out ablation experiments to demonstrate the effectiveness and contribution of each core component introduced in CONCAVEQ on 3s\_vs\_5z scenario in SMAC. As shown in Fig.\ref{exp_ablation} we consider verifying the effect of (1) concave mixing network by replacing it with 4-layer monotonic network as \textit{CONCAVEQ\_linear},  (2) iterative action selection which removes the iterative procedure as \textit{CONCAVEQ\_no\_iter}, (3) soft policy network which removes the soft policy network as \textit{CONCAVEQ\_no\_policy},  (4) disable both iterative action selection and soft policy network as \textit{CONCAVEQ\_disabled}, (5) further remove the central $Q^*$ as \textit{CONCAVE\_no\_$Q^*$}. The results show that CONCAVEQ overperforms these ablated versions. \textit{CONCAVEQ\_no\_iter} has lower rewards than \textit{CONCAVEQ} due to a lack of iterative action selection which is unfavorable to finding the optimal actions. Removing the soft policy network also leads to a performance drop due to a lack of a proper action selection scheme during execution.  The reward of \textit{CONCAVEQ\_linear} grows slowly at first as linear layers have weaker representation ability than concave mixing networks.  \textit{CONCAVEQ\_no\_$Q^*$} suffers from the most significant performance drop after most core components are removed from the original design. Such results validate how each component is crucial for achieving performance through experiments. 

\setlength\intextsep{0pt}

%% file: 7Conclusion.tex
\section{Conclusions}
In this paper, we propose ConcaveQ, a novel non-monotonic value function factorization approach, which goes beyond monotonic mixing functions that are known to have limited representation expressiveness. ConcaveQ employs neural network representations of concave mixing functions. 
An iterative action selection scheme is developed to obtain optimal action during training, while factorized policies using local networks are leveraged to support fully decentralized execution. 
We evaluate the proposed ConcaveQ on predator-prey and StarCraft II tasks. Empirical results demonstrate substantial improvement of ConcaveQ over state-of-the-art MARL algorithms with monotonic factorization.

\clearpage
\section{Acknowledgement}
This work was partially supported by the National Natural Science Foundation
of China (NSFC) under Grant 62122042 and in part by Major Basic Research Program of
Shandong Provincial Natural Science Foundation under Grant ZR2022ZD02.

%% file: 8Appendix.tex
\onecolumn

\begin{center}
 \Large\textbf{Appendix \\
}
\end{center}

\setcounter{section}{0}
\setcounter{theorem}{0}
\setcounter{equation}{0}
\setcounter{prop}{0}

\section{Organization of the Appendix}
In Section A.1, we present the theorems and their corresponding proofs concerning ConcaveQ. Section A.2 describes the experimental setup in detail. We show Predator Prey Environment in Section A.2.1 and show SMAC Environment in Section A.2.2.  We demonstrate implementation details in Section A.3.

\section{A.1 \ \ Proof of Theorems} \label{subsec:Theorems}
In this section, we show the proof of all the Theorems of this work.

\begin{theorem}
 When $n \geq log_{|S|}(2|A|\cdot log_2|A|) + 1$ and the optimal action choices in $Q_{jt}$ are uniformly distributed, for any monotonic factorization, and for some constant $\delta\in(0,1)$, we have

 \begin{equation}
     \frac{E(|S_{mono}|)}{|S|^n} \leq \frac{e+1}{|A|}\cdot \delta^{n-1}
 \end{equation}

\end{theorem}

\noindent\textit{Proof. \ }The key idea of the upper bound proof of monotonic representation limitations is to convert the problem into a classic max-load problem.

\noindent \textit{\textbf{Step1}: Formulate as max-load bin-ball problem.} For each agent $i$ and state $\mathbf{s}$, we consider the optimal action of $Q_{mono}$ as \textit{ball $i$}. Thus, $Q_{jt}$ and $Q_{mono}$ have the same optimal action for  agent $i$ if ball $i$ is placed in the bin corresponding to the optimal action of $Q_{jt}$. 


Let $X_i$ denotes that ball $i$ is in bin $i$, that is,  $Q_{jt}$ and $Q_{mono}$ have the same optimal action for  agent $i$, then we have: 
\begin{equation}
    E(|S_{mono}|) = \Sigma_sP(X_1)\cdot P(X_2|X_1) \cdot...\cdot P(X_n|X_{n-1}...X_1) \nonumber
\label{S_mono}
\end{equation}

Define $Y_i$ as the load of bin $i$, that is, the state space such that $Q_{jt}$ and $Q_{mono}$ have the same optimal action for agents except for agent $i$.
Note that the global maximum of $Q_{jt}$ is uniformly distributed over different states, we then analyze $P(X_1)$:
\begin{equation}
P(X_1) =  \frac{E(\Sigma_{s_2...s_n}X_1)}{|S|^{n-1}}
\leq \frac{E(max_{i}Y_i)}{|S|^{n-1}}
\label{P_X_1}
\end{equation}

\noindent \noindent\textit{\textbf{Step2}: Analyze the probability distribution of the load.} 
By the union bound, we have:
\begin{equation} 
P(max_{i}Y_i) \geq \frac{e \cdot {|S|^{n-1}}}{|A|} \leq \Sigma_iP(Y_i \geq \frac{e \cdot {|S|^{n-1}}}{|A|})
\label{P_maxY_union}
\end{equation}
By the Chernoff bound, for any $\alpha$, we have,
\begin{equation} 
P((max_{i}Y_i) \geq (1+\alpha)EY_i) \leq ( \frac{e^\alpha }{(1+\alpha)^{(1+\alpha)}})^{EY_i}
\label{P_maxY_union}
\end{equation}
Let us assume that $\alpha \geq e-1 $. Then, we have,
\begin{equation} 
\begin{aligned}
\frac{e^\alpha }{(1+\alpha)^{(1+\alpha)}} = &\frac{1}{1+\alpha}\cdot (\frac{e}{1+\alpha})^\alpha \\
\leq &\frac{1}{1+\alpha} \\
\leq &\frac{1}{2}
\end{aligned}
\label{P_maxY_union}
\end{equation}

\begin{equation} 
\begin{aligned}
\frac{e^\alpha }{(1+\alpha)^{(1+\alpha)}} = \frac{1}{1+\alpha}\cdot (\frac{e}{1+\alpha})^\alpha 
\leq \frac{1}{1+\alpha} 
\leq \frac{1}{2}
\end{aligned}
\label{P_maxY_union}
\end{equation}
Since  $EY_i \geq \frac{|S|^{n-1}}{|A|}$,  we have,

\begin{equation} 
P(max_{i}Y_i \geq \frac{e\dot|S|^{n-1}}{|A|}) \leq ( \frac{1}{2})^\frac{|S|^{n-1}}{|A|}
\label{P_maxY_union}
\end{equation}

\begin{equation}
\begin{aligned}
E(max_{i}Y_i) =  &E(max_{i}Y_i|Y_i \geq \frac{e\dot|S|^{n-1}}{|A|}) \cdot P(Y_i \geq \frac{e\dot|S|^{n-1}}{|A|}) \\ 
+ &E(max_{i}Y_i|Y_i < \frac{e\dot|S|^{n-1}}{|A|}) \cdot P(Y_i < \frac{e\dot|S|^{n-1}}{|A|}) \\
\leq &\frac{|A|\cdot|S|^{n-1}}{2^\frac{|S|^{n-1}}{|A|}} + \frac{e\cdot |S|^{n-1}}{|A|} 
\end{aligned}
\label{P_maxY_union}
\end{equation}
When $\frac{1}{2^\frac{|S|^{n-1}}{|A|}} \leq \frac{1}{|A|^2}$ or $n \geq log_{|S|}(2|A|\cdot log_2|A|) + 1$, we have,
\begin{equation}
E(max_{i}Y_i) \leq \frac{(e+1)|S|^{n-1}}{|A|}
\label{E_maxY}
\end{equation}
where $|A| \geq 1$ is the size of action space. 
Applying Eq. (\ref{E_maxY}) to Eq. (\ref{P_X_1}), we have:
\begin{equation}
P(X_1) \leq \frac{e+1}{|A|}
\end{equation}
Using the same argument repeatedly for agents $i=2,\ldots,n$, we can choose $\delta = min_i(P(X_i|X_{i-1}...X_1))$ and $0 < \delta < 1$. Plugging these inequalities into $ E(|S_{mono}|) $, we have, 
\begin{equation}
\begin{aligned}
E(|S_{mono}|) &\leq \Sigma_s\frac{e+1}{|A|}\cdot \delta^{n-1}  \\
&= \frac{|S|^n\cdot(e+1)}{|A|}\cdot \delta^{n-1}
\end{aligned}
\end{equation}
\begin{equation}
\frac{E(|S_{mono}|)}{|S|^n} \leq \frac{e+1}{|A|}\cdot \delta^{n-1} 
\end{equation}

{\begin{prop}
For any state $s$, there is always a concave function $Q_{c}(s, \textbf{u})$ that recovers the global maximum of  $Q_{jt}(s, \textbf{u})$ with the same optimal action.
\end{prop}

\noindent\textit{Proof. \ } For any state $s$, let $f_1(\textbf{u})$ = $Q_{jt}(, \textbf{u})$ and $f_2(\textbf{u})$ = -$Q_{jt}(, \textbf{u})$. According to Fenchel–Moreau theorem \cite{convex_conjugate}, the biconjugate function of $f_2(\textbf{u})$ is a convex function $g(\textbf{u}) = f_2^{**}(\textbf{u})$ and $h(\textbf{u}) = -f_2^{**}(\textbf{u})$  is a concave function. $g(\textbf{u})$ and $h(\textbf{u})$ satisfies:
 \begin{equation}
    g(\textbf{u}) \leq f_2(\textbf{u}), 
\end{equation}
 \begin{equation}
    -g(\textbf{u}) \geq -f_2(\textbf{u}), 
\end{equation}
 \begin{equation}
    h(\textbf{u}) \geq f_1(\textbf{u}).
\end{equation}
Suppose the optimal joint action of $f_1(\textbf{u})$ and $h(\textbf{u})$ are $\textbf{u}_{f_1}^*$ and $\textbf{u}_h^*$ respectively,  if we shift $h(\textbf{u})$ by $(-\textbf{u}_{f_1}^* + \textbf{u}_h^*)$, the shifted concave function has the same optimal action as $f_1(\textbf{u})$. In other words, for any state $s$, there is always a concave function $h(\textbf{u}-(\textbf{u}_{f_1}^* + \textbf{u}_h^*)$ that has the same optimal action as $Q_{jt}(s, \textbf{u})$. 
}

\begin{theorem}
The mixing network $f$ is concave in $x$ provided that all $W^{(z)}_{0:k-1}$ are non-negative, and all functions $a_i$ are convex and non-decreasing.
\end{theorem}

\noindent\textit{Proof. \ } 
Since linear functions are concave and convex, $z_1$ is a convex function of $x$. Note that t non-negative sums of convex functions are also convex and that the composition of a convex and convex non-decreasing function is also convex, $z_{i+1}, i = 1, ..., k-2$ is also a convex function of $x$. Following the fact that the negative
of a convex function is a concave, $z_k$ is a concave function of $x$. 

\section{A2 \ \ Algorithm Analysis}

\subsection{Iterative Action Selection}
~\\
\begin{algorithm} 
	\caption{Iterative action selection} 
    \label{alg_action_selction} 
	\begin{algorithmic}
        \STATE Initialize action and action-value function with greedy action search, and let $\mathbf{u}_{opt} = \mathbf{u}_{init}$, $Q_{tot} = Q_{init}$
        \FOR{$agent_i = 0$ to $max\_agents$}
        \FOR{ $action_i = 0$ to $max\_actions$}
        \STATE replace $(\boldsymbol{u}_{opt})^{(agent_i)}$ with $action_i$ and get $\boldsymbol{u}_{current}$        
        \IF {$Q(s, \boldsymbol{u}_{current}) > Q_{tot}$}
        \STATE $Q_{tot} = Q(s, \boldsymbol{u}_{current})$
        \STATE $\boldsymbol{u}_{opt} = \boldsymbol{u}_{current}$
        
        \ENDIF
        \ENDFOR
        \ENDFOR \\
        Return $\boldsymbol{u}_{opt}$
	\end{algorithmic} 
\end{algorithm}
\subsection{Factorized soft policy iteration}
Our factorized soft policy is based on Boltzmann exploration policy iteration. Previous works have demonstrated that within the MARL domain, Boltzmann exploration policy iteration is proven to enhance the policy as well as converge with a certain number of iterations. In this context, the Boltzmann policy iteration is  defined as:

\begin{equation}
J(\pi)=\sum_{t} \mathbb{E}\left[r\left(\mathbf{s}_{t}, \mathbf{u}_{t}\right)+\alpha \mathcal{H}\left(\pi\left(\cdot | \mathbf{s}_{t}\right)\right)\right]
\end{equation}

 The gradient for factorized soft policy can be given via:

\begin{equation}
\begin{array}{l}
\begin{aligned}
\mathcal{L}_{\pi}(\pi) &=\mathbb{E}_{\mathcal{D}}\left[\alpha \log \boldsymbol{\pi}\left(\boldsymbol{u}_{t} | \boldsymbol{\tau}_{t}\right)-Q^{\pi}_{tot}\left(\boldsymbol{s_{t}}, \boldsymbol{\tau_{t}}, \boldsymbol{u}_{t}\right)\right] \\
&= -q^{\pi}\left(\boldsymbol{s}_{t}, \mathbb{E}_{\pi^{i}}\left[q^{i}\left(\tau_{t}^{i}, u_{t}^{i}\right)-\alpha \log \pi^{i}\left(u_{t}^{i} | \tau_{t}^{i}\right)\right]\right)
\end{aligned}
\end{array}
\end{equation}

Let  $q^{\pi}$ be the operator of a one-layer mixing network with no activation functions at the end whose parameters are generated from the hyper-network with input $\boldsymbol{s}_{t}$, then
\begin{equation}
\begin{aligned}
& q^{\pi}(\boldsymbol{s}_{t}, \boldsymbol{q}(\tau_{t}, a_{t}) - \boldsymbol{\alpha} \log \boldsymbol{\pi(a_{t}| \tau_{t}})) 
\\  \\
& = \sum_{i}[ w^{i}(\boldsymbol{s}) \mathbb{E}_{\pi} [q^{i}(\tau_{t}^{i}, a_{t}^{i})]  - \sum_{i}[ w^{i}(\boldsymbol{s})\alpha^{i} \log {\pi^{i}}({a}_{t} |{\tau}_{t})] + b(\boldsymbol{s}) 
\label{q_pi_1}
\end{aligned}
\end{equation}

where $ w^{i}(s)$ and $b^{i}(s)$ are the corresponding weights and biases of $q^{\pi}$ conditioned on $\boldsymbol{s}$).
Analyze  the first  and second item in Eq. \ref{q_pi_1} respectively, and we have,
\begin{equation}
q^{\pi}(\boldsymbol{s}_{t}, \boldsymbol{q}(\tau_{t}, a_{t})) = \sum_{i}[ w^{i}(\boldsymbol{s}) \mathbb{E}_{\pi} [q^{i}(\tau_{t}^{i}, a_{t}^{i})] + b(\boldsymbol{s}) = \mathbb{E}_{\pi} [Q_{tot}(\boldsymbol{\tau}, \boldsymbol{a} ; \boldsymbol{\theta})] 
\label{q_pi_2}
\end{equation}

\begin{equation}
\mathbb{E}_{\pi} [Q_{tot}(\boldsymbol{\tau}, \boldsymbol{a} ; \boldsymbol{\theta})] = \sum_{\boldsymbol{a}} {\pi^{i}}({a}_{t}^{i} |{\tau}_{t}^{i}) \mathbb{E}_{\pi} [Q_{tot}(\boldsymbol{\tau}, \boldsymbol{a} ; \boldsymbol{\theta})] 
\label{q_pi_3}
\end{equation}

Applying Eq. \ref{q_pi_2} and Eq. \ref{q_pi_1} to Eq. \ref{q_pi_1}, we have,
\begin{equation}
\begin{aligned}
q^{\pi}(\boldsymbol{s}_{t}, \boldsymbol{q}(\tau_{t}, a_{t}) - \boldsymbol{\alpha} \log \boldsymbol{\pi(a_{t}| \tau_{t}})) &= \mathbb{E}_{\pi} [Q_{tot}(\boldsymbol{\tau}, \boldsymbol{a} ; \boldsymbol{\theta})] - \sum_{i}[ w^{i}(\boldsymbol{s})\mathbb{E}_{\pi} [\alpha^{i} \log {\pi^{i}}({a}_{t} |{\tau}_{t})]] \\
  &= \mathbb{E}_{\pi} [Q_{tot}(\boldsymbol{\tau}, \boldsymbol{a} ; \boldsymbol{\theta})] -\sum_{i} \mathbb{E}_{\pi} [\boldsymbol{\alpha} \log {\pi^{i}}({a}_{t}^{i} |{\tau}_{t}^{i})] \\
& \text{\hspace{3em}(let $\alpha^{i} =  \frac{\boldsymbol\alpha}{w^{i}(\boldsymbol{s})}$)} \\
&= \mathbb{E}_{\pi} [Q_{tot}(\boldsymbol{\tau}, \boldsymbol{a} ; \boldsymbol{\theta})] - \sum_{i}\sum_{\pi} [\boldsymbol{\alpha}\pi^{i} ({a}_{t}^{i} |{\tau}_{t}^{i})\log {\pi^{i}}({a}_{t}^{i} |{\tau}_{t}^{i})] \\ 
& = \mathbb{E}_{\pi} [Q_{tot}(\boldsymbol{\tau}, \boldsymbol{a} ; \boldsymbol{\theta})] - \sum_{\pi} \boldsymbol{\alpha} \log \boldsymbol{\pi}(\boldsymbol{a}_{t} |\boldsymbol{\tau}_{t}) \\
& \text{\hspace{3em}(Assume $\boldsymbol\pi = \prod \pi^{i}$, then $\sum_{i}\sum_{\pi} [\boldsymbol{\alpha}\pi^{i} ({a}_{t}^{i} |{\tau}_{t}^{i})\log {\pi^{i}}({a}_{t}^{i} |{\tau}_{t}^{i})] = \sum_{\pi} \boldsymbol{\alpha} \log \boldsymbol{\pi}(\boldsymbol{a}_{t} |\boldsymbol{\tau}_{t})$}) \\
& = \mathbb{E}_{\pi}[ Q^{\pi}_{tot}(\boldsymbol{s_{t}}, \boldsymbol{\tau_{t}}, \boldsymbol{a}_{t})-\boldsymbol\alpha \log \boldsymbol{\pi}(\boldsymbol{a}_{t} | \boldsymbol{\tau}_{t})]\\
\end{aligned}
\end{equation}

We leverage the derivation above to demonstrate that utilizing  $\boldsymbol{q^\pi}(\tau_{t}, a_{t}) - \boldsymbol{\alpha} \log \boldsymbol{\pi(a_{t}| \tau_{t}})$ directly as input for the mixing network can serve as soft-actor-critic policy update policy in a value factorization approach. This holds when applying a single-layer mixing network without an activation function. Although this condition provides insights into the proposed design, using a ReLU activation function allows this expression to serve as a lower bound for optimization.

\section{A2 \ \ Environment Details}

\begin{algorithm}
    \caption{pseudocode for training CONCAVEQ  }
    \begin{algorithmic}
        \FOR{$k = 0 $ to $max\_train\_steps$}                    
            \STATE {Initialize the environment, mixing network  $Q^*, Q_{tot}$, critic network $q$, policy network $\pi$}
            \STATE {Initialize the Replay buffer $\mathcal{D}$}
            \FOR{$t = 0 $ to $max\_episode\_limits$}    
              \STATE Execute joint action $\mathbf{a}$, observe reward $r$, and observation $\boldsymbol{\tau}$, next state $s_{t+1}$
              
              \STATE Store ($\boldsymbol{a}$, r, $\boldsymbol{\tau}$, $\boldsymbol{\tau^{'}}$) pair in replay buffer  $\mathcal{D}$
            \ENDFOR
        \FOR{t = 1 to N}
            \STATE Sample trajectory minibatch $\mathcal{B}$ from $\mathcal{D}$
            \STATE Take action $a_{i}\sim \pi_{i}$
            
            \STATE Calculate Loss 
            \STATE \quad\quad\quad $\mathcal{L}(\theta) = \mathcal{L}_{\pi} +  \mathcal{L}_{\hat{Q^{*}}} + \mathcal{L}_{{\text{ConcaveQ}}}$
            
            \STATE Update parameters of the critic network and mixing networks
            \STATE \quad\quad\quad $\boldsymbol{\theta(q, Q^*, Q_{tot})} \gets \beta\nabla\mathcal{L}(\theta)$ 
            \STATE Update policy network 
            \STATE \quad\quad\quad $\boldsymbol{\theta}(\pi) \gets \beta\nabla\mathcal{L}(\pi)$

            \STATE  Update the temperature parameter
            \STATE \quad\quad\quad  $\alpha \gets \beta \nabla\alpha$ 
            \IF{$t$ mod T = 0}
                \STATE Update target networks: $\boldsymbol{\theta^{\prime}} \gets \boldsymbol{\theta} $
            \ENDIF
        \ENDFOR
        \ENDFOR
    \end{algorithmic}
\end{algorithm}
For evaluation, we adopt state-of-the-art baselines that are closely related to our work and the most recent, such as   QMIX (baseline for value-based factorization methods), WQMIX (uses weighted projections to enhance representation ability), RESQ \cite{ResQ} (which is the most advanced value-based method), PAC \cite{pac}, QPLEX \cite{QPLEX} and  FOP \cite{fop} (SOTA actor-critic based method).   Our code implementation is available on GitHub.


\subsection{A2.1 \ Predator Prey}\label{predator}

 Predator-prey Task is widely adopted to simulate a  partially observable environment on a  grid-world setting to study relative overgeneralization problem \cite{deep_coori_graph}. In this setup, a grid of size 10 × 10 is utilized, housing 8 agents tasked with capturing 8 prey. Agents possess the options of moving in any of the four cardinal directions, staying stationary, or attempting to ensnare adjacent prey. 
 Impossible actions, such as moving into an already occupied target position or catching when no adjacent prey exists, are designated as unavailable. Upon the execution of a catch action by two neighboring agents, a prey is successfully captured and both the prey and capturing agents are eliminated from the grid. The observation afforded to an agent comprises a $5 × 5$ sub-grid with the agent at the center, encompassing two distinct channels: one indicating the presence of agents and the other denoting the presence of prey.  An episode concludes either when all agents are removed from the grid or after 200 timesteps have transpired. The act of capturing prey garners a reward of $r$ = 10, whereas unsuccessful solo attempts incur a negative penalty of $p$. This study undertakes four sets of experiments, each with varying $p$ values: $p$ = 0, $p$ = -0.5, $p$ = -1.5, and $p$ = -2.

\subsection{A2.2 \ SMAC}\label{smac}
We conducted experiments on StarCraft II micromanagement following the instructions in \cite{smac} with open-source code including QMIX \cite{QMIX}, WQMIX \cite{WQMIX},  RESQ \cite{ResQ}, PAC \cite{pac}, FOP \cite{fop}, and QPLEX \cite{QPLEX}. We consider combat scenarios where the enemy units are controlled by the built-in AI in StarCraft II, while the friendly units are in the control of agents trained by MARL algorithms. The built-in AI difficulties cover a large range: Very Easy, Easy, Medium, Hard, Very Hard, and Insane, ranging from 0 to 7. We carry out the experiments with the highest difficulty for built-in AI: difficulty = 7 (Insane). Depending on the specific scenarios(maps), the units of the enemy and friendly can be either symmetric or asymmetric.  Each agent selects one action from discrete action space, like no-op, move[direction], attack[enemy\_id], and stop at each time step. Dead units are restricted to choosing the no-op action. Eliminating an enemy unit yields a reward of 10 while achieving victory by eliminating all enemy units results in a reward of 200. The global state information is solely available to the centralized critic. The map scenarios used in experiments are shown in Table \ref{tab:my-table} We train each baseline algorithm with 3 distinct random seeds. Evaluation is performed every 10,000 training steps with 32 testing episodes for the main results.  For ablation results, evaluation is carried out with three random seeds.  The experimentation is conducted on a Nvidia GeForce RTX 3080 Ti workstation, and on average, it takes approximately 2 hours to finish one run for the 3s5z map on the SMAC environment.

\begin{table*}[ht!]
\centering
\resizebox{\textwidth}{!}{%
\begin{tabular}{@{}ccc@{}}
\toprule
map          & Ally Units                          & Enemy Units                         \\ \midrule
1c3s5z       & 1 Colossus, 3 Stalkers \& 5 Zealots & 1 Colossus, 3 Stalkers \& 5 Zealots \\
3m           & 3 Marines                           & 3 Marines                           \\
3s5z         & 3 Stalkers \& 5 Zealots             & 3 Stalkers \& 5 Zealots             \\
8m           & 8 Marines                           & 8 Marines                           \\ \midrule
3s\_vs\_5z   & 3 Stalkers                          & 5 Zealots                           \\
5m\_vs\_6m   & 5 Marines                           & 6 Marines                           \\
6h\_vs\_8z   & 6 Hydralisks                        & 8 Zealots                           \\
27m\_vs\_30m & 27 Marines                          & 30 Marines                          \\
Corridor     & 6 Zealots                           & 24 Zerglings                        \\
MMM2         & 1 Medivac, 2 Marauders \& 7 Marines & 1 Medivac, 3 Marauders \& 8 Marines \\\bottomrule
\end{tabular}%
}
\caption{Concise overview of SMAC Map scenarios employed in experimental settings}
\label{tab:my-table}
\end{table*}

\section{A3 \ \ Implementation Details}

We utilize one set of hyper-parameters across each environment, abstaining from tailored adjustments for specific maps.  In the absence of explicit specification, uniform configurations for general hyperparameters, such as learning rate, are maintained across all algorithms, while algorithm-specific hyperparameters remain at their default values. For action selection, we employ an epsilon-greedy strategy with an annealing schedule from an initial $\epsilon$ value of 0.995, gradually decreasing to $\epsilon$ = 0.05 over the course of 100,000 training steps in a linear fashion.
We use epsilon greedy for action selection with annealing from $\epsilon$ = 0.995 decreasing to $\epsilon$ = 0.05 in 100000 training steps linearly. 
Batch size set as $bs$ = 128, with a replay buffer capacity of 10,000 instances. And the learning rate is denoted as $lr$ = 0.001.
$\beta$ value is set to 0.001, considering the comparable dimensions of $o_i$ and $\hat{u}^*_i$, thus mitigating the need for excessive compression.
Weighting functions employ weights $w$ = 0.5 and the temporal Difference lambda parameter $\lambda$ set at 0.6.
Initial entropy term logarithmically set to log$\alpha$ = -0.07, its learning rate designated as $lr_{\alpha}$ = 0.0003.
Target network update interval established at every 200 episodes. And
algorithm performance evaluation conducted over 32 episodes every 1000 training steps.


%% file: main.bbl
\begin{thebibliography}{29}
\providecommand{\natexlab}[1]{#1}

\bibitem[{B{\"o}hmer, Kurin, and Whiteson(2020)}]{deep_coori_graph}
B{\"o}hmer, W.; Kurin, V.; and Whiteson, S. 2020.
\newblock Deep coordination graphs.
\newblock In \emph{PMLR}.

\bibitem[{Borwein and Lewis(2006)}]{convex_conjugate}
Borwein, J.; and Lewis, A. 2006.
\newblock \emph{Convex Analysis and Nonlinear Optimization: Theory and Examples}.
\newblock Springer.

\bibitem[{Boyd and Vandenberghe(2004)}]{convex_optimization}
Boyd, S.; and Vandenberghe, L. 2004.
\newblock \emph{Convex Optimization}.
\newblock Cambridge university press.

\bibitem[{Chen et~al.(2022{\natexlab{a}})Chen, Chen, Lan, and Aggarwal}]{chen2022multi}
Chen, J.; Chen, J.; Lan, T.; and Aggarwal, V. 2022{\natexlab{a}}.
\newblock Multi-agent covering option discovery based on kronecker product of factor graphs.
\newblock \emph{IEEE Transactions on Artificial Intelligence}.

\bibitem[{Chen et~al.(2022{\natexlab{b}})Chen, Chen, Lan, and Aggarwal}]{NEURIPS2022_c40d1e40}
Chen, J.; Chen, J.; Lan, T.; and Aggarwal, V. 2022{\natexlab{b}}.
\newblock Scalable Multi-agent Covering Option Discovery based on Kronecker Graphs.
\newblock In \emph{Advances in Neural Information Processing Systems}, volume~35, 30406--30418. Curran Associates, Inc.

\bibitem[{Chen and Lan(2023)}]{chen2023minimizing}
Chen, J.; and Lan, T. 2023.
\newblock Minimizing return gaps with discrete communications in decentralized pomdp.
\newblock \emph{arXiv preprint arXiv:2308.03358}.

\bibitem[{Czumaj(2004)}]{balls_into_bins}
Czumaj, A. 2004.
\newblock Lecture notes in Approximation and Randomized Algorithms.
\newblock \url{https://www.ic.unicamp.br/~celio/peer2peer/math/czumaj-balls-into-bins.pdf}.

\bibitem[{Foerster et~al.(2018)Foerster, Farquhar, Afouras, Nardelli, and Whiteson}]{coma}
Foerster, J.~N.; Farquhar, G.; Afouras, T.; Nardelli, N.; and Whiteson, S. 2018.
\newblock Counterfactual Multi-Agent Policy Gradients.
\newblock In \emph{AAAI}.

\bibitem[{Gogineni et~al.(2023)Gogineni, Mei, Lan, Wei, and Venkataramani}]{gogineni2023accmer}
Gogineni, K.; Mei, Y.; Lan, T.; Wei, P.; and Venkataramani, G. 2023.
\newblock Accmer: Accelerating multi-agent experience replay with cache locality-aware prioritization.
\newblock In \emph{2023 IEEE 34th International Conference on Application-specific Systems, Architectures and Processors (ASAP)}, 205--212. IEEE.

\bibitem[{Hu et~al.(2021)Hu, Jiang, Harding, Wu, and Liao}]{marl}
Hu, J.; Jiang, S.; Harding, S.~A.; Wu, H.; and Liao, S.-w. 2021.
\newblock Rethinking the implementation tricks and monotonicity constraint in cooperative multi-agent reinforcement learning.
\newblock In \emph{ICLR}.

\bibitem[{Iqbal and Sha(2019)}]{maac}
Iqbal, S.; and Sha, F. 2019.
\newblock Actor-Attention-Critic for Multi-Agent Reinforcement Learning.
\newblock In \emph{ICML}.

\bibitem[{Karush(1939)}]{concaveopt}
Karush, W. 1939.
\newblock \emph{Minima of Functions of Several Variables with Inequalities as Side Conditions}.
\newblock Master's thesis, Department of Mathematics, University of Chicago.

\bibitem[{Lowe et~al.(2017)Lowe, Wu, Tamar, Harb, Abbeel, and Mordatch}]{maddpg}
Lowe, R.; Wu, Y.; Tamar, A.; Harb, J.; Abbeel, P.; and Mordatch, I. 2017.
\newblock Multi-Agent Actor-Critic for Mixed Cooperative-Competitive Environments.
\newblock In \emph{NeuralIPS}.

\bibitem[{Mei, Zhou, and Lan(2023)}]{mei2023remix}
Mei, Y.; Zhou, H.; and Lan, T. 2023.
\newblock ReMIX: Regret Minimization for Monotonic Value Function Factorization in Multiagent Reinforcement Learning.
\newblock \emph{arXiv preprint arXiv:2302.05593}.

\bibitem[{Mei et~al.(2023)Mei, Zhou, Lan, Venkataramani, and Wei}]{mei2022mac}
Mei, Y.; Zhou, H.; Lan, T.; Venkataramani, G.; and Wei, P. 2023.
\newblock MAC-PO: Multi-Agent Experience Replay via Collective Priority Optimization.
\newblock In \emph{Proceedings of the 2023 International Conference on Autonomous Agents and Multiagent Systems}, AAMAS '23, 466–475. International Foundation for Autonomous Agents and Multiagent Systems.

\bibitem[{Rashid et~al.(2020)Rashid, Farquhar, Peng, and Whiteson}]{WQMIX}
Rashid, T.; Farquhar, G.; Peng, B.; and Whiteson, S. 2020.
\newblock Weighted {QMIX:} Expanding Monotonic Value Function Factorisation for Deep Multi-Agent Reinforcement Learning.
\newblock In \emph{NeuralIPS}.

\bibitem[{Rashid et~al.(2018)Rashid, Samvelyan, de~Witt, Farquhar, Foerster, and Whiteson}]{QMIX}
Rashid, T.; Samvelyan, M.; de~Witt, C.~S.; Farquhar, G.; Foerster, J.~N.; and Whiteson, S. 2018.
\newblock {QMIX:} Monotonic Value Function Factorisation for Deep Multi-Agent Reinforcement Learning.
\newblock In \emph{{ICML}}.

\bibitem[{Samvelyan et~al.(2019{\natexlab{a}})Samvelyan, Rashid, De~Witt, Farquhar, Nardelli, Rudner, Hung, Torr, Foerster, and Whiteson}]{samvelyan2019starcraft}
Samvelyan, M.; Rashid, T.; De~Witt, C.~S.; Farquhar, G.; Nardelli, N.; Rudner, T.~G.; Hung, C.-M.; Torr, P.~H.; Foerster, J.; and Whiteson, S. 2019{\natexlab{a}}.
\newblock The starcraft multi-agent challenge.
\newblock \emph{arXiv preprint arXiv:1902.04043}.

\bibitem[{Samvelyan et~al.(2019{\natexlab{b}})Samvelyan, Rashid, De~Witt, Farquhar, Nardelli, Rudner, Hung, Torr, Foerster, and Whiteson}]{smac}
Samvelyan, M.; Rashid, T.; De~Witt, C.~S.; Farquhar, G.; Nardelli, N.; Rudner, T.~G.; Hung, C.-M.; Torr, P.~H.; Foerster, J.; and Whiteson, S. 2019{\natexlab{b}}.
\newblock The starcraft multi-agent challenge.
\newblock \emph{arXiv preprint arXiv:1902.04043}.

\bibitem[{Shen et~al.(2022)Shen, Qiu, Liu, Liu, Fu, Liu, and Wang}]{ResQ}
Shen, S.; Qiu, M.; Liu, J.; Liu, W.; Fu, Y.; Liu, X.; and Wang, C. 2022.
\newblock ResQ: A Residual Q Function-based Approach for Multi-Agent Reinforcement Learning Value Factorization.
\newblock In \emph{{NeurIPS}}.

\bibitem[{Son et~al.(2019)Son, Kim, Kang, Hostallero, and Yi}]{QTRAN}
Son, K.; Kim, D.; Kang, W.~J.; Hostallero, D.; and Yi, Y. 2019.
\newblock {QTRAN:} Learning to Factorize with Transformation for Cooperative Multi-Agent Reinforcement Learning.
\newblock In \emph{ICML}.

\bibitem[{Su, Adams, and Beling(2021)}]{VDAC}
Su, J.; Adams, S.; and Beling, P. 2021.
\newblock Value-decomposition multi-agent actor-critics.
\newblock In \emph{AAAI}.

\bibitem[{Sunehag et~al.(2018)Sunehag, Lever, Gruslys, Czarnecki, Zambaldi, Jaderberg, Lanctot, Sonnerat, Leibo, Tuyls, and Graepel}]{VDN}
Sunehag, P.; Lever, G.; Gruslys, A.; Czarnecki, W.~M.; Zambaldi, V.~F.; Jaderberg, M.; Lanctot, M.; Sonnerat, N.; Leibo, J.~Z.; Tuyls, K.; and Graepel, T. 2018.
\newblock Value-Decomposition Networks For Cooperative Multi-Agent Learning Based On Team Reward.
\newblock In \emph{{AAMAS}}.

\bibitem[{Wang et~al.(2021{\natexlab{a}})Wang, Ren, Liu, Yu, and Zhang}]{QPLEX}
Wang, J.; Ren, Z.; Liu, T.; Yu, Y.; and Zhang, C. 2021{\natexlab{a}}.
\newblock {QPLEX:} Duplex Dueling Multi-Agent Q-Learning.
\newblock In \emph{ICLR}.

\bibitem[{Wang et~al.(2021{\natexlab{b}})Wang, Han, Wang, Dong, and Zhang}]{dop}
Wang, Y.; Han, B.; Wang, T.; Dong, H.; and Zhang, C. 2021{\natexlab{b}}.
\newblock DOP: Off-Policy Multi-Agent Decomposed Policy Gradients.
\newblock In \emph{ICLR}.

\bibitem[{Yang et~al.(2020)Yang, Hao, Liao, Shao, Chen, Liu, and Tang}]{qatten}
Yang, Y.; Hao, J.; Liao, B.; Shao, K.; Chen, G.; Liu, W.; and Tang, H. 2020.
\newblock Qatten: A general framework for cooperative multiagent reinforcement learning.
\newblock \emph{arXiv preprint arXiv:2002.03939}.

\bibitem[{Zhang et~al.(2021)Zhang, Li, Wang, Xie, and Lu}]{fop}
Zhang, T.; Li, Y.; Wang, C.; Xie, G.; and Lu, Z. 2021.
\newblock FOP: Factorizing Optimal Joint Policy of Maximum-Entropy Multi-Agent Reinforcement Learning.
\newblock In \emph{ICML}.

\bibitem[{Zhou, Lan, and Aggarwal(2022)}]{pac}
Zhou, H.; Lan, T.; and Aggarwal, V. 2022.
\newblock PAC: Assisted Value Factorization with Counterfactual Predictions in Multi-Agent Reinforcement Learning.
\newblock In \emph{NeurlIPS}.

\bibitem[{Zhou, Lan, and Aggarwal(2023)}]{zhou2022value}
Zhou, H.; Lan, T.; and Aggarwal, V. 2023.
\newblock Value Functions Factorization With Latent State Information Sharing in Decentralized Multi-Agent Policy Gradients.
\newblock \emph{IEEE Transactions on Emerging Topics in Computational Intelligence}, 7(5): 1351--1361.

\end{thebibliography}
